\documentclass[10pt,onecolumn]{IEEEtran}

\usepackage{hyperref}
\usepackage{multirow}
\usepackage{emp}
\usepackage{graphicx}
\usepackage[T1]{fontenc}
\usepackage{booktabs}
\usepackage{caption}
\usepackage{subcaption}
\usepackage{amsmath}
\usepackage{amsfonts}
\usepackage{amssymb}

\newcommand{\specialcell}[2][c]{%
  \begin{tabular}[#1]{@{}c@{}}#2\end{tabular}}

\ifCLASSINFOpdf

\else

\fi

\begin{document}
\title{\Large\bfseries Reality Mining with Mobile Big Data: \\ Understanding the Impact of Network Structure on Propagation Dynamics}
%\title{When the Vehicle Routing Problem Met Big Data Analytics}
%\title{Big Data Analytics based Networking for Energy-Efficient Vehicle Routing}
\author{
\IEEEauthorblockN{Yuanfang Chen\dag, Noel Crespi\dag, Gyu Myoung Lee\ddag} \\
\IEEEauthorblockA{
\dag Institut Mines-T\'el\'ecom, T\'el\'ecom SudParis, France \\
\dag Department of Computer Science, Universit\'e Pierre et Marie CURIE, France \\
\ddag Liverpool John Moores University, Liverpool, UK \\
%\ddag Dalian University of Technology, China lvlin\_george@mail.dlut.edu.cn\\
Email: yuanfang\_chen@ieee.org, noel.crespi@mines-telecom.fr, G.M.Lee@ljmu.ac.uk}
}

\maketitle

%=====================================================================================================================================================
%Big Data Analytics based Networking for Energy-Efficient Vehicle Routing
\begin{abstract}
Information and epidemic propagation dynamics in complex networks is truly important to discover and control terrorist attack and disease spread.  How to track, recognize and model such dynamics is a big challenge.  With the popularity of intellectualization and the rapid development of Internet of Things (IoT), massive mobile data is automatically collected by millions of wireless devices (e.g., smart phone and tablet).  In this article, as a typical use case, the impact of network structure on epidemic propagation dynamics is investigated by using the mobile data collected from the smart phones carried by the volunteers of Ebola outbreak areas.  On this basis, we propose a model to recognize the dynamic structure of a network.  Then, we introduce and discuss the open issues and future work for developing the proposed recognition model.
\end{abstract}

\IEEEpeerreviewmaketitle
%On this basis, tracking, recognizing and modelling the propagation dynamics is possible.
%=======================================================================================================================================================

\section{Introduction}
Information and epidemic propagation dynamics~\cite{rodriguez2014uncovering, wen2013modeling, theodorakopoulos2013selfish, fu2013propagation} has been extensively studied by network-enabled science, e.g., graph theory, network theory and probability theory.  When information and epidemic propagation is modelled over networks, it is usual to assume that the propagation has the same probability over links.  Even if different links have respective propagation probabilities, such modelling is not enough to reflect the real propagation pattern in the physical world.  As the important feature of networks, network structure needs to be considered~\cite{guo2011activation}.  For example, the patterns of propagation are different in a scale-free network and a network with heavy-tailed distribution.

It is an open issue whether the structure of complex networks underlies the propagation dynamics of information and epidemic~\cite{pei2013spreading}.  Despite a lack of direct experimental evidence supporting such ``structure-propagation'' hypothesis, a number of theoretical studies have shown that the topological structure of complex networks (scale-free and small-world topologies) leads to markedly different propagation dynamics compared with the predicted by standard propagation models.  For example, in the literature~\cite{small2007scale}, Michael Small~\emph{et al.} examine the global spatio-temporal distribution of avian influenza cases in both wild and domestic birds, and they find that the cases and the links between the cases during an outbreak form a scale-free network.  It means that such avian influenza outbreak will continue to propagate even with a vanishingly small propagation rate.  In contrast, with standard mathematical models of disease propagation~\cite{brauer2012mathematical}, the propagation of this avian influenza has been controlled and even has halted.  It makes the best time to vaccination be missed, and the risk of second outbreak be increased.

Based on the above description, understanding the impact of network structure on propagation dynamics is very important, and recognizing the dynamic structure of a network is a gap for the previous studies of propagation dynamics.

\begin{figure}[!ht]
  \centering
  \includegraphics[width=2.5in]{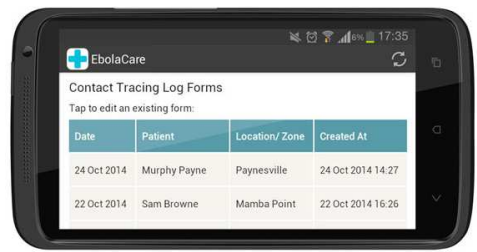}\\
  \caption{Contact Tracing module of Ebola Care~\cite{ebolacare}.  It can track everyone who came in direct contact with a sick Ebola patient.  The collected data by this APP is shared with the WHO (World Health Organization), who is using information from hundreds of aid organizations to make big strategic decisions.}
  \label{fig:ebola_app}
\end{figure}

This article reviews the advance of propagation dynamics, and then as a typical use case we investigate the impact of network structure on epidemic propagation dynamics based on the mobile data collected from the GPS-enabled wireless devices carried by volunteers (Fig.~\ref{fig:ebola_app} illustrates an example).  On this basis, we propose a model to recognize the dynamic structure of a network.  Finally, open issues and future work are provided and discussed for developing the proposed recognition model.

%=======================================================================================================================================================

\section{Advance of Propagation Dynamics}
It is important to understand the propagation processes arising over the networks with different structures, for example, knowledge mining: how a behaviour on a network with a special structure to impact the nodes of the network, and such understanding is helpful to model the behaviour as well.  In recent years, there is an increasing effort to study propagation dynamics based on a variety of complex networks.  Recent achievements can be divided into two categories based on different types of networks:
\begin{itemize}
  \item Propagation dynamics on social networks~\cite{lappas2010finding}.  On such networks, information is the main research target.  Exponential and power-law models that reflect network structure have been widely used to model the dynamics of information propagation.
  \item Propagation dynamics on contact networks~\cite{vazquez2013using}.  A contact network describes the real relationships among individuals/ecosystems in the physical world.  Based on the real relationships from the physical world, the propagation dynamics on contact networks is different from the propagation dynamics on social networks.  With the development of the IoT (Internet of Things) and the help of various sensors and wireless devices, some researchers have paid their attention to this propagation dynamics, and have obtained some achievements in: (i) the propagation of infectious diseases, and (ii) the propagation of contaminants.  Analyzing and studying the dynamics of propagation among individuals/ecosystems can help us to understand and control the dynamic behaviours on these real networks.
\end{itemize}

As an important aspect of propagation dynamics, the theoretical studies on the ``structure-propagation'' hypothesis are classified into two classes: information-related and epidemic-related propagation dynamics on respective complex networks.

As the important recent achievements in information-related propagation dynamics~\cite{rodriguez2014uncovering}, Jure Leskovec~\emph{et al.} obtain three interesting observations, along with tracking information propagation among media sites and blogs: (i) the information pathways for general recurrent topics are more stable across time than for on-going news events.  It means that the former has a more stable network structure.  (ii) Clusters of news media sites and blogs often emerge and vanish in a matter of days for on-going news events.  From this observation, we can acquire that hub nodes (clusters) are existent in an information propagation network.  As the key element to reflect network structure, the clusters are dynamically changed over time, and different information propagation networks are with different clusters.  And (iii) major events, for example, large-scale civil unrest such as Libyan civil war and Syrian uprising, increase the number of information pathways among blogs, and also increase the network centrality of blogs and social media sites.  Moreover, as a kind of harmful information propagation, malware/virus propagation dynamics is also an important research issue.  Guanhua Yan~\emph{et al.} investigate the characteristics of malware propagation in online social networks based on a data set collected from a real-world location-based online social network, which includes not only the social graph formed by its users but also the activity events of these users.  And further, they study the impact of social structure on malware propagation dynamics in online social networks.  This achievement is to deeply understand the impact of online social network structure on malware propagation dynamics.

For the recent achievement in epidemic-related propagation dynamics~\cite{kim2014estimating}, Louis Kim~\emph{et al.} propose a parameter estimation method by learning network characteristics and disease dynamics.  And this method is applied to the data collected during the 2009 H1N1 epidemic, and on this basis, they find the outbreak network is best fitted into a scale-free network.  This finding implies that random vaccination alone will not efficiently halt the propagation of influenza, and instead vaccination should be based on understanding the propagation dynamics of epidemic with exploiting the special structure of network.

Moreover, as an important aspect of epidemic-related propagation dynamics, quantifying and predicting disease dynamics during epidemics is very important to public health in allocating public health resources and in responding to public health events.  The infected number $R$ can be used to quantify the disease dynamics during an epidemic.  For studying the quantized disease dynamics, a wide range of methods have been proposed to estimate or predict the parameter $R$ based on the assumptions of network structure, e.g., the contact networks for the spread of disease are best described as having exponential degree distributions.

However, network structure is time-varying along with the propagation of information and epidemic on the network.  It is necessary to recognize the dynamic structure of such network.  For example, for improving the accuracy of estimating and predicting for $R$ during an epidemic in a network, the dynamic structure of the network needs to be mined.

%=======================================================================================================================================================

\section{Impact of Network Structure on Propagation Dynamics}
As a typical use case, the impact of network structure on epidemic propagation dynamics is investigated based on the contact network of Ebola outbreak in 2014.

\begin{figure}[!ht]
  \centering
  \includegraphics[width=2.5in]{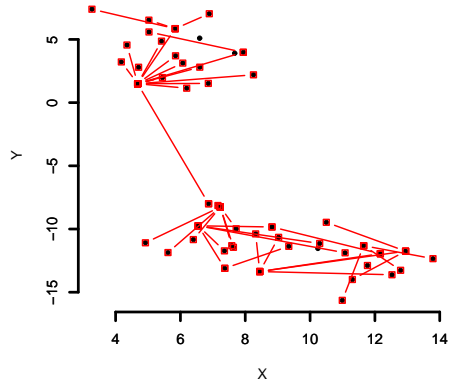}\\
  \caption{An example of the contact network during an epidemic.  This example displays 50 cases and their relationships (contact).  X and Y denote the relative locations of cases.  These cases come from three typical countries and seven regions of the Ebola outbreak in 2014.  Three countries are: Guinea, Nigeria and Liberia.  Seven regions are: Gueckedou, Macenta, Kissidougou, Conakry, Monrovia, Lagos and Port Harcourt.  The black nodes of this network are cases (suspected and confirmed) and if there is an edge between two nodes, it means that there is contact between the individuals of the two cases.}
  \label{fig:ebola_network}
\end{figure}

With the wireless communication devices held by volunteers of epidemic areas, the volunteers report new cases (confirmed and suspected cases), corresponding locations, and relationships between these cases, and then, these reported cases with corresponding locations can be used to build the contact network (an example is shown in Fig.~\ref{fig:ebola_network}).  During an epidemic, the network is time-varying along with the propagation of an infectious disease, with the order of time stamps of reports.  The contact network can be modelled as a dynamic graph $G_{t}$.  The weight $w_{\{i,j\}}$ is the transmission probability ($p_{\{i,j\}}$) of a disease from vertex $i$ to vertex $j$ (on the corresponding edge $e_{\{i,j\}}$).

%In total there are 942 nodes in this network.
\subsection{Outbreak Data}
The outbreak data of Ebola in West Africa from March 2014 is used as real surveillance data to analyze the impact of network structure on propagation dynamics.

As a latest outbreak of disease, until February 15, 2015, Ebola Virus Disease (it is commonly known as ``Ebola'') has killed 9380 people, and the total cases have reached 23253.  Researchers generally believe that from a 2-year-old boy of Guinea to his mother, sister and grandmother (a contact network), Ebola rapidly spreads in West Africa from March 2014.

The reported Ebola cases with time series and location information are collected by the World Health Organization (WHO), as well as the ministries of health of epidemic countries.  And in this study, we select part of data from three typical outbreak countries, Guinea, Nigeria and Liberia.  Guinea is the source of this outbreak and is with relatively high quantity of confirmed cases (2727, as of February 15, 2015), and Nigeria is far away from the source of the outbreak, and is with relatively low quantity of confirmed cases (19, as of February 15, 2015), and Liberia is close to the source of the outbreak, and is with high quantity of confirmed cases (3149, as of February 15, 2015).  And seven regions of these three countries are: Gueckedou of Guinea, Macenta of Guinea, Kissidougou of Guinea, Conakry of Guinea, Monrovia of Liberia, Lagos of Nigeria, and Port Harcourt of Nigeria.  And these variables are included in the outbreak data: (i) Case ID.  A unique number indicates a case. (ii) Source ID.  A source id indicates the source of infection for a case.  (iii) Date.  It is the date that a case is reported.  (iv) Location.  It indicates the coordinates (longitude and latitude) of a reported case.

\subsection{Results and Analysis}
\label{sec:results}
To evaluate the impact of network structure on disease dynamics, the basic and important structural knowledge of networks, degree distribution, is measured for the contact network that is studied in this article.
%three typical properties are very important to measure the network structure: degree distribution, clustering coefficient (transitivity) and degree correlation (assortativity).

In a network, the degree of a node is basic structural knowledge, and it indicates the number of adjacent edges of the node.  The degree distribution is the probability distribution of degrees over the network.  It gives the overall structural information of the network.  For a real-world network, there are complex relationships among nodes.  The degree distribution is helpful to characterize and model a real-world network.  On this basis, the structural knowledge of a complex network can be acquired and formulated.  The formulated knowledge can be used to analyze and solve network-related problems.

In this article, we analyze the degree distribution of the contact network acquired from the collected Ebola outbreak data, by conducting the maximum-likelihood fitting to fit the calculated degree distribution of the network into exponential, normal, poisson and power-law distributions, and calculating and comparing the estimated standard deviations and the estimated variance-covariance matrices of these fittings.

\begin{figure*}[!ht]
  \centering
  \begin{subfigure}[t]{0.4\textwidth}
    \includegraphics[width=2.5in]{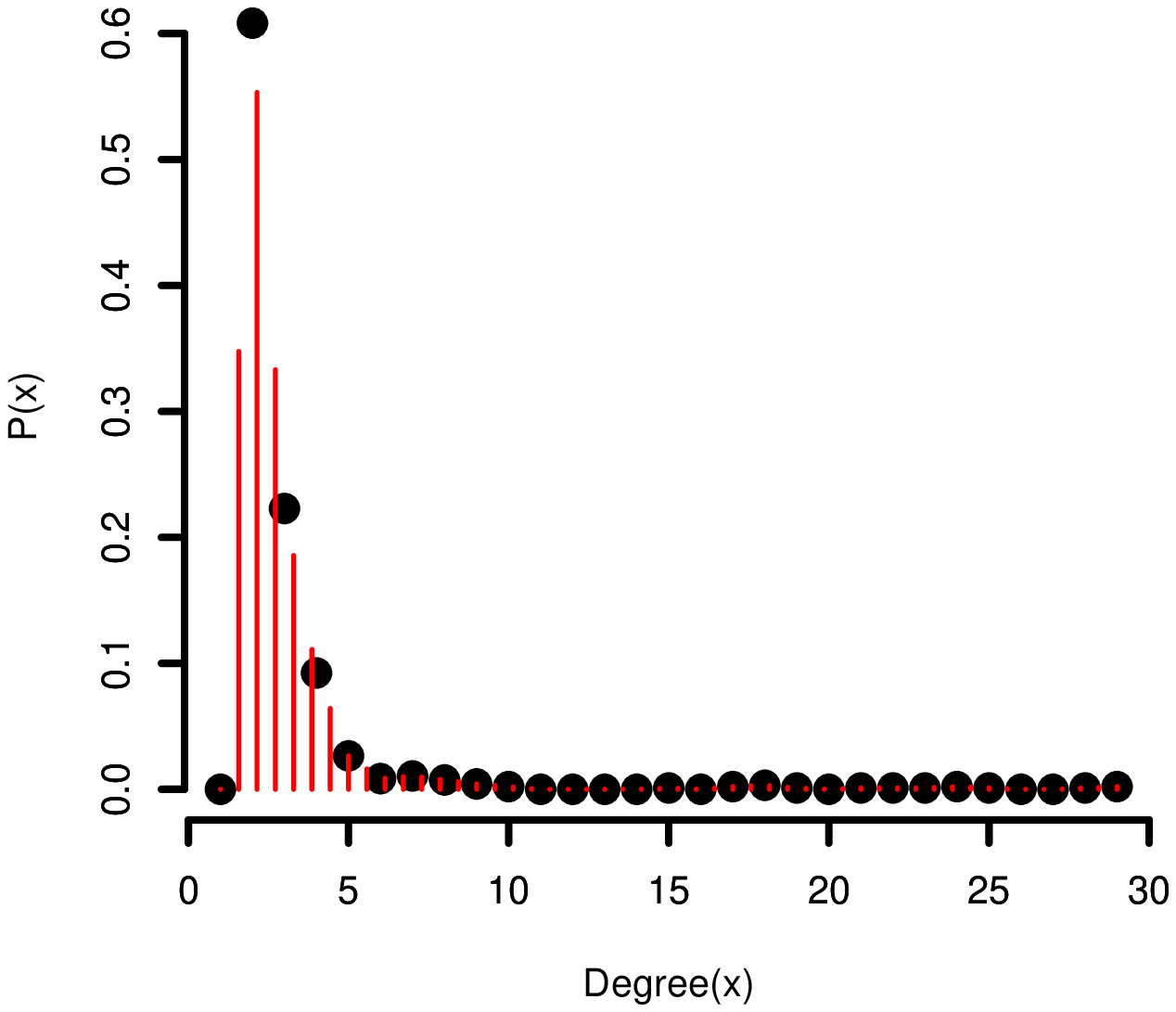}
    \caption{Degree distribution of the contact network built by Ebola outbreak data.  There are 942 nodes and 938 edges in this network.  The black spots are the probability distribution of nodes' degrees.}
    \label{fig:degree_distribution_human}
  \end{subfigure}
  \qquad
  \begin{subfigure}[t]{0.4\textwidth}
    \includegraphics[width=2.5in]{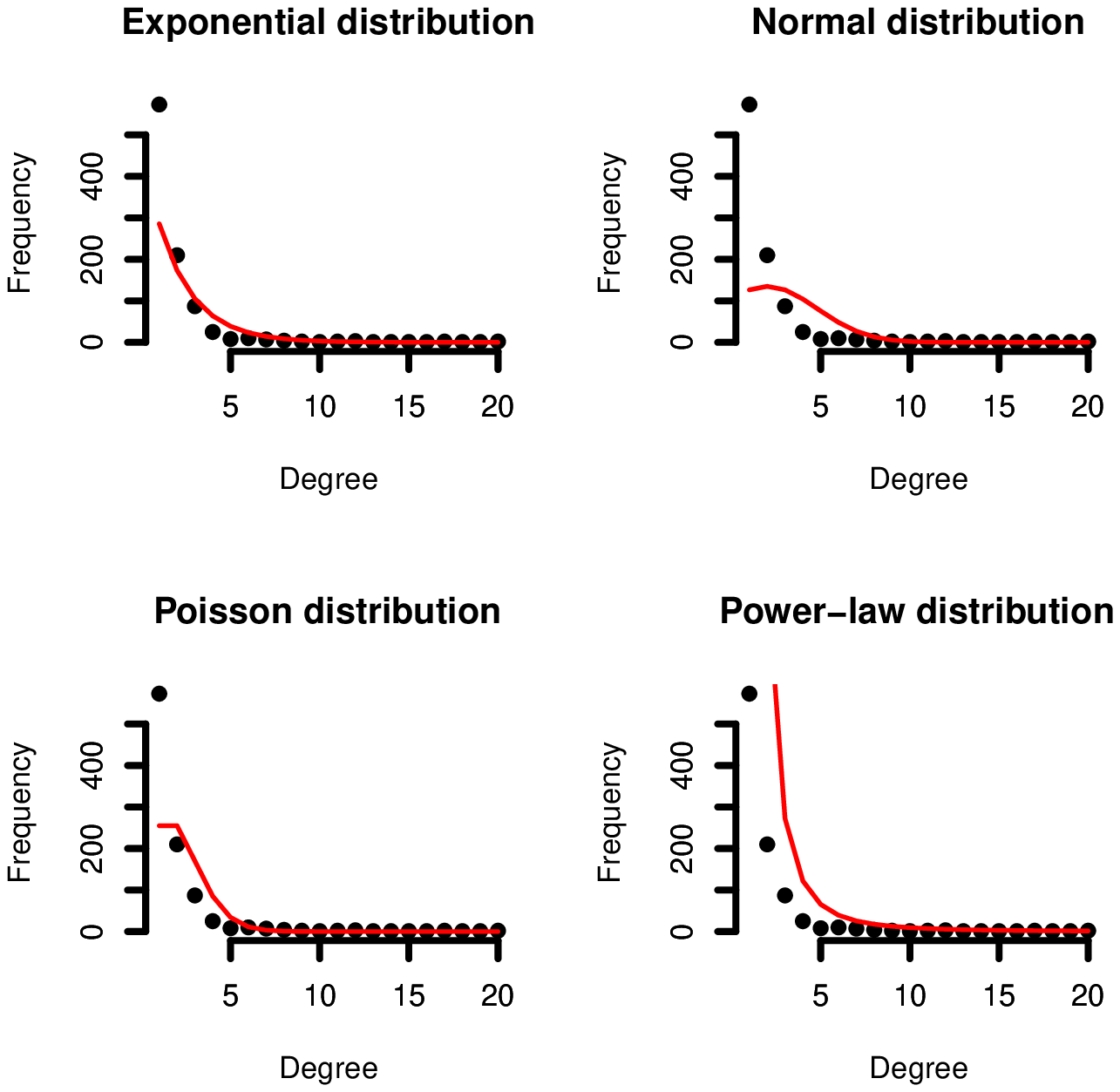}
    \caption{Maximum-likelihood fitting of degree distributions.  The degree distribution of contact network is fitted into exponential, normal, poisson and power-law distributions with maximum-likelihood fitting.  The black spots indicate the probability distribution of nodes' degrees, and the red lines are the corresponding fittings for exponential, normal, poisson and power-law distributions.}
    \label{fig:degree_fitting_human}
  \end{subfigure}
  \caption{Degree distribution and maximum-likelihood fitting for the contact network of Ebola outbreak}
  \label{fig:human_degree_distribution_and_fittings}
\end{figure*}

Figure~\ref{fig:degree_distribution_human} illustrates the degree distribution of the contact network built by Ebola outbreak data.  On Fig.~\ref{fig:degree_distribution_human} basis, for analyzing the degree distribution of contact network, maximum-likelihood fitting is conducted to fit the degree distribution into exponential, normal, poisson and power-law distributions, and then the estimated standard deviations and the estimated variance-covariance matrices of these fittings are measured to quantify ``how many difference between two different distributions''.  The results of fittings are illustrated in Fig.~\ref{fig:degree_fitting_human}.

In Fig.~\ref{fig:human_degree_distribution_and_fittings}, the results show that the degree distribution of contact network is approximate to the exponential distribution with $\lambda=0.50159915$.

For different distributions, based on the maximum-likelihood fitting, the results of parameter estimation are listed as follows: (i) the rate parameter $\lambda=0.50159915$ for the exponential distribution, (ii) $\mu=1.99362380$ and $\sigma=2.77914691$ for the normal distribution, (iii) $\lambda=1.9936238$ for the poisson distribution, and (iv) $x_{min}=2$ and $\alpha=2.803973$ for the power-law distribution.

Table~\ref{tab:deviations_matrices_human} shows the estimated standard deviations and the estimated variance-covariance matrices of these fittings.
\begin{table*}[!ht]
\centering
\caption{Estimated standard deviations and estimated variance-covariance matrices for different fittings}
\begin{tabular}{c|c|c}
  \hline
  Distribution & Standard deviation & Variance-covariance matrix \\ \hline
  Exponential & $\lambda$ (rate parameter): 0.01635166 & $\begin{array}{c|c}
                                & rate~parameter \\ \hline
                                rate~parameter & 2.673769e-04
                              \end{array}$
  \\ \hline
  Normal & \specialcell{$\mu$ (mean): 0.09059760, \\ $\sigma$ (standard deviation (sd)): 0.06406218} & $\begin{array}{c|c|c}
                                & mean & sd \\ \hline
                                mean & 0.008207925 & 0.000000000 \\ \hline
                                sd & 0.000000000 & 0.004103963
                              \end{array}$
  \\ \hline
  Poisson & $\lambda$ (lambda): 0.0460285 & $\begin{array}{c|c}
                                & lambda \\ \hline
                                lambda & 0.002118623
                              \end{array}$
  \\ \hline
  Power-law & $x_{min}+\alpha$: 0.03831463 & NULL \\
  \hline
\end{tabular}
\label{tab:deviations_matrices_human}
\end{table*}

Comparing the estimated standard deviations and estimated variance-covariance matrices listed in Tab.~\ref{tab:deviations_matrices_human}, the minimum standard deviation for these fittings is $0.01635166$.  This minimum standard deviation is corresponding to the exponential distribution with the rate parameter $\lambda=0.50159915$.

\begin{figure*}[!ht]
  \centering
  \begin{subfigure}[t]{0.4\textwidth}
    \includegraphics[width=2.5in]{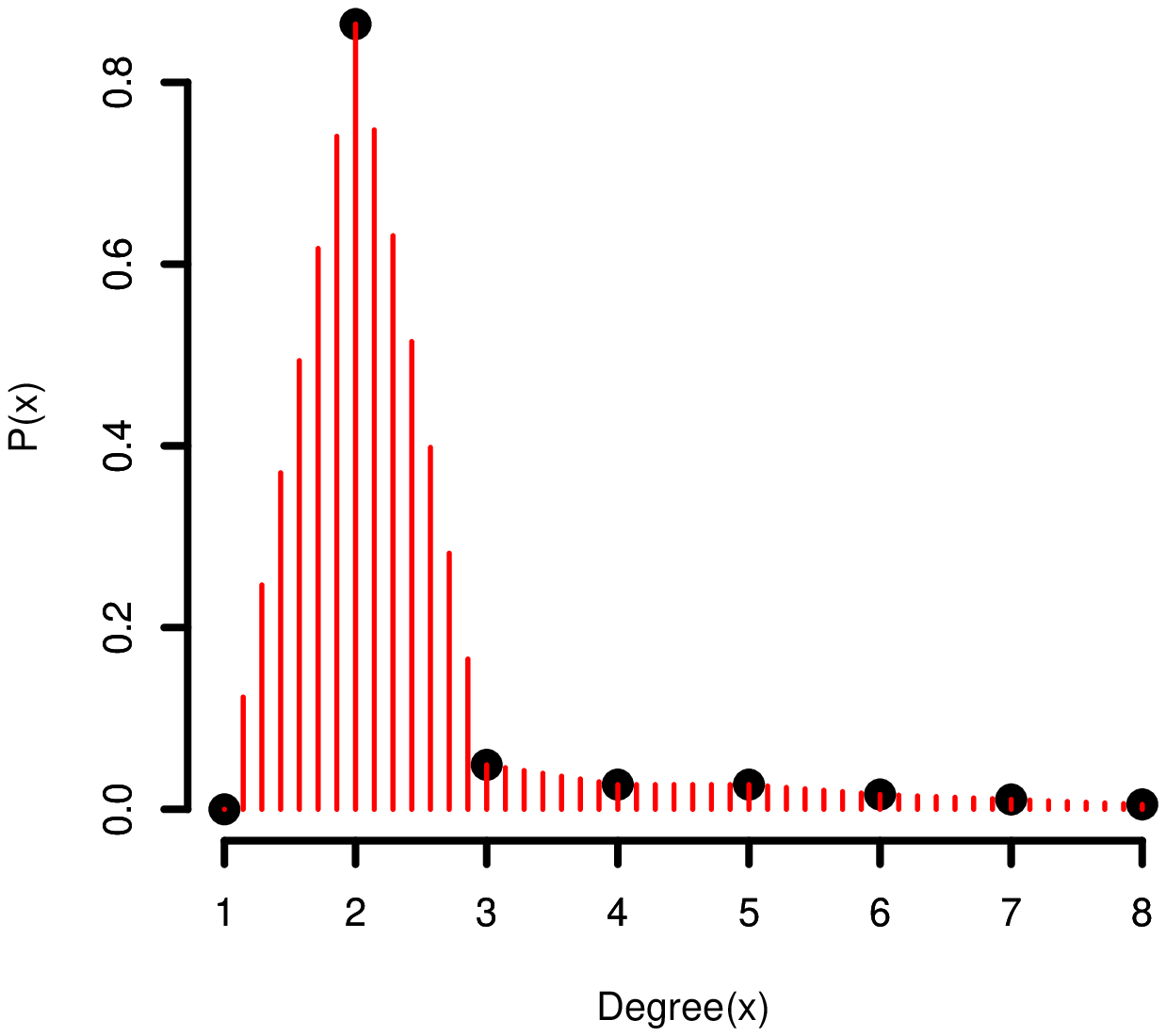}
    \caption{Degree distribution for the subnetwork of contact network.  There are $96$ time periods of August 26th, 2014 in this subnetwork.  The black spots are the probability distribution of nodes' degrees.}
    \label{fig:degree_distribution_human_subnetwork}
  \end{subfigure}
  \qquad
  \begin{subfigure}[t]{0.4\textwidth}
    \includegraphics[width=2.5in]{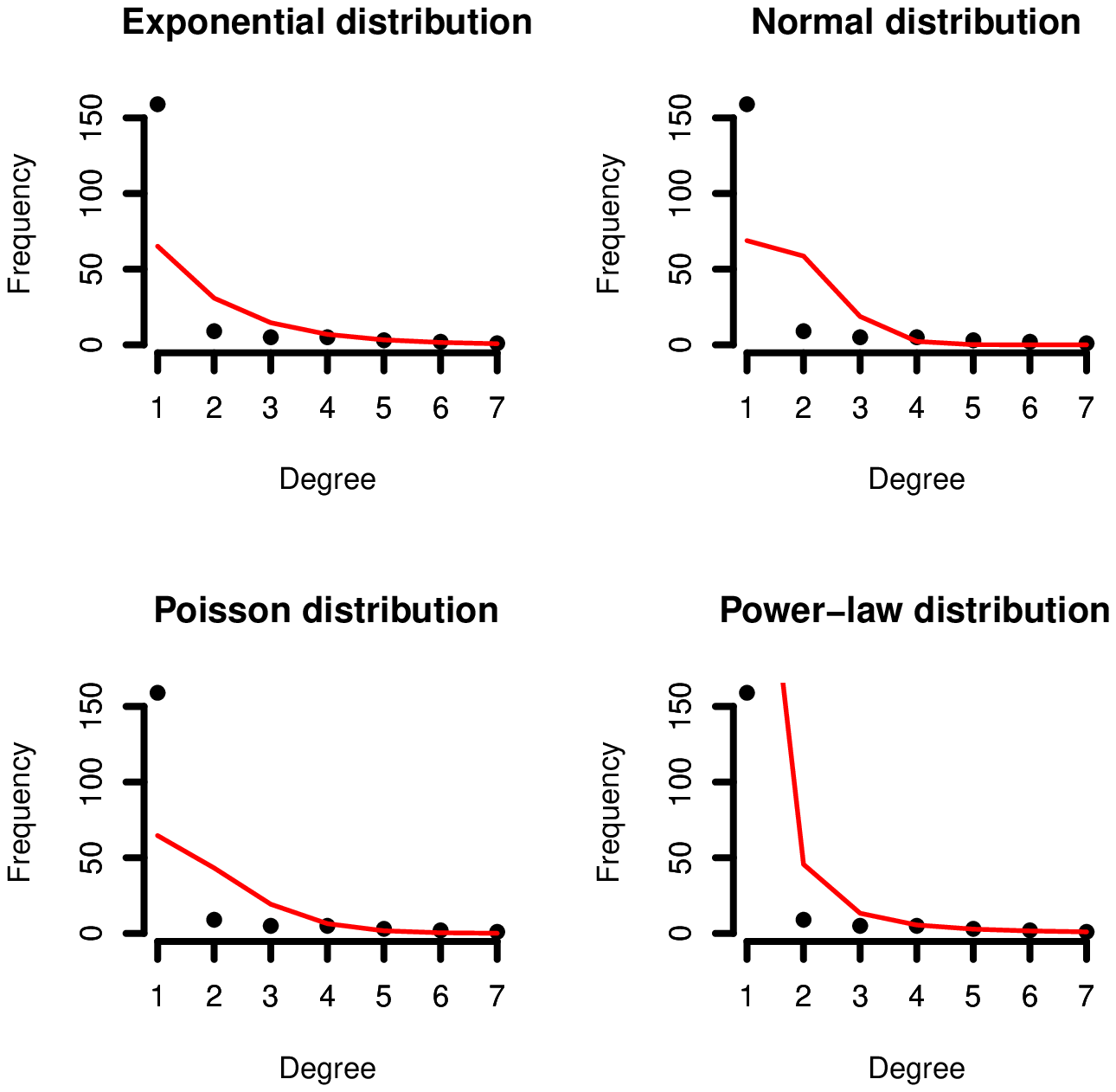}
    \caption{Maximum-likelihood fitting of degree distributions.  The degree distribution of the subnetwork is fitted into exponential, normal, poisson and power-law distributions by maximum-likelihood fitting.  The black spots indicate the probability distribution of nodes' degrees, and the red lines are the corresponding fittings for exponential, normal, poisson and power-law distributions.}
    \label{fig:degree_fitting_human_subnetwork}
  \end{subfigure}
  \caption{Degree distribution and maximum-likelihood fitting for the subnetwork of contact network}
  \label{fig:human_degree_distribution_and_fittings_subnetwork}
\end{figure*}

However, based on the description of the network that is studied in this article, the contact network is time-varying along with the propagation of infectious disease.  As an example, the analysis results of the subnetwork that is with $96$ time periods of August 26th, 2014, are shown in Fig.~\ref{fig:human_degree_distribution_and_fittings_subnetwork} and Tab.~\ref{tab:deviations_matrices_human_subnetwork}.

By the maximum-likelihood fitting for the subnetwork, the results of parameter estimation for different distributions are: (i) the rate parameter $\lambda=0.74796748$ for the exponential distribution, (ii) $\mu=1.33695652$ and $\sigma=1.00841216$ for the normal distribution, (iii) $\lambda=1.33695652$ for the poisson distribution, and (iv) $x_{min}=1$ and $\alpha=3.041947$ for the power-law distribution.

Table~\ref{tab:deviations_matrices_human_subnetwork} shows the estimated standard deviations and the estimated variance-covariance matrices of the fittings for the subnetwork.
\begin{table*}[!ht]
\centering
\caption{Estimated standard deviations and estimated variance-covariance matrices for different fittings}
\begin{tabular}{c|c|c}
  \hline
  Distribution & Standard deviation & Variance-covariance matrix \\ \hline
  Exponential & $\lambda$ (rate parameter): 0.05514089 & $\begin{array}{c|c}
                                & rate~parameter \\ \hline
                                rate~parameter & 0.003040518
                              \end{array}$
  \\ \hline
  Normal & \specialcell{$\mu$ (mean): 0.07434113, \\ $\sigma$ (standard deviation (sd)): 0.05256712} & $\begin{array}{c|c|c}
                                & mean & sd \\ \hline
                                mean & 0.005526604 & 0.000000000 \\ \hline
                                sd & 0.000000000 & 0.002763302
                              \end{array}$
  \\ \hline
  Poisson & $\lambda$ (lambda): 0.08524123 & $\begin{array}{c|c}
                                & lambda \\ \hline
                                lambda & 0.007266068
                              \end{array}$
  \\ \hline
  Power-law & $x_{min}+\alpha$: 0.02865438 & NULL \\
  \hline
\end{tabular}
\label{tab:deviations_matrices_human_subnetwork}
\end{table*}

From the fitting results for the subnetwork of August 26th, 2014, which are listed in Tab.~\ref{tab:deviations_matrices_human_subnetwork}, the degree distribution of the subnetwork is approximate to the power-law distribution with $x_{min}=1$ and $\alpha=3.041947$.

Based on the above detailed analyses on the structure of networks, these two facts can be observed: (i) epidemic propagation shows high pertinency with network structure.  It means that along with the propagation of disease during an epidemic, the extending of contact network is to follow a certain pattern.  That is, network structure impacts the dynamics of epidemic propagation: the propagation dynamics is different on different networks with different structures.  (ii) Network structure is time-varying during an epidemic (it makes a dynamic network).

%=======================================================================================================================================================

\section{Recognition Model of Network Structure}
How to recognize the dynamic structure of a contact network is a valuable research issue.  It is important to quantify and predict the propagation dynamics during an epidemic.  If the quantification and prediction can be achieved, it will be helpful to accurately allocate public health resources and respond to public health events.

\begin{figure}[!ht]
  \centering
  \includegraphics[width=3.5in]{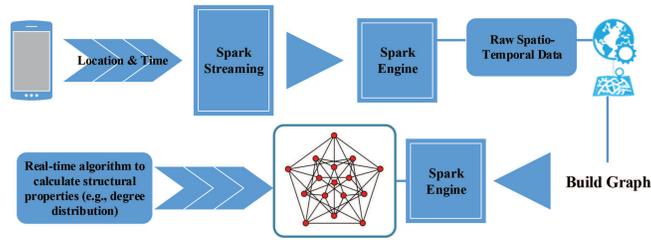}\\
  \caption{Recognition model workflow}
  \label{fig:model_workflow}
\end{figure}

Based on the analytical ability of Apache Spark~\cite{spark} on Streaming Data and Graph, we propose the recognition model of network structure.  The workflow of this model is illustrated in Fig.~\ref{fig:model_workflow}.

In this model, the input stream is composed of the spatio-temporal data, GPS locations and times, from smart phones.  The GPS location of a case is associated with the physical location where the case is found and reported.  On this basis, the cases and their relationships from within a period of time (by analyzing the time stamp of each recorded case), are used to build a graph.  In this graph, the distance between two vertices can be calculated based on the coordinates of these two vertices.  The graph is time-varying: the vertex and edge set are changed over time along with the propagation of epidemic.

By the spark engine with a real-time algorithm and the processing of Spark Streaming~\footnote{Spark Streaming provides a language-integrated API to stream processing.  It makes the processing of streaming data be easy as processing batch data.}, such dynamic change of graph can be followed, and structural properties (e.g., degree distribution) and maximum-likelihood fitting can be calculated.  With the calculated properties and fitting, the propagation dynamics of an epidemic can be quantized and even the trend of propagation can be predicted.  A prediction for an epidemic is based on the actual movements of cases, and with obtained special information of network structure, the prediction will be more effective.

%=======================================================================================================================================================

\section{Open Issues and Future Work}
The above proposed recognition model is based on Spark to process and analyze streaming data and the data-based graph.  For developing this model, these open issues need to be studied as future work:
\begin{itemize}
  \item Building a graph based on spatio-temporal data.  Once the spatio-temporal data is persistently input, Spark can process it by Spark Streaming, and further through the graph algorithms of Spark Engine, a dynamic graph is built.  From ``data'' to ``graph'', there is an important process, ``transformation''.  For a certain problem, such process is with specific requirements, for example, a suitable period of time needs to be confirmed to build the graph that is used to denote the real-time contact network of epidemic propagation.
  \item Real-time algorithm design based on the framework of Spark.  Spark is designed for big data analytics.  It means that Spark is a fast and general engine for large-scale data processing.  In this engine, the processing is based on parallelization technology, so traditional serial algorithms are not suitably used in here.
  \item Following the dynamic change of graph.  It is necessary to design a real-time algorithm to recognize the structure of dynamic graph and calculate the structural properties of the graph.
\end{itemize}

%=======================================================================================================================================================

\section{Conclusion}
By analyzing the impact of network structure on epidemic propagation dynamics, this article has proposed a model to recognize the dynamic structure of network.  This model is based on the ability of Spark to process and analyze streaming data and the data-based graph.  The model can calculate: (i) the structural properties of dynamic network, and (ii) the maximum-likelihood fitting for these structural properties.  By this calculation, the structure of network can be recognized in real time.  Specifically, from the point of view of the future work, we have introduced and discussed the open issues for developing this model.

%=======================================================================================================================================================

\bibliographystyle{IEEEtran}
\bibliography{mag}

%=======================================================================================================================================================

\end{document}